# Adaptive Cohort Size Determination Method for Bayesian Optimal Interval Phase I/II Design to Shorten Clinical Trial Duration

Masahiro Kojima[1,2]

[1]Biometrics Department, R&D Division, Kyowa Kirin Co., Ltd., Tokyo, Japan.

[2]Research Center for Medical and Health Data Science, The Institute of Statistical Mathematics, Tokyo, Japan.

**Running title**: Adaptive Cohort Size Determination Method

**Corresponding author**

Name: Masahiro Kojima, Ph.D.

Address: Biometrics Department, R&D Division, Kyowa Kirin Co., Ltd.

Otemachi Financial City Grand Cube, 1-9-2 Otemachi, Chiyoda-ku, Tokyo 100-0004, Japan.

Tel: +81-3-5205-7200

FAX: +81- 3-5205-7182

Email: masahiro.kojima.tk@kyowakirin.com

**Financial Support**: Not applicable.

**Word count**: 2757

**The number of figures**: 4

**The number of tables**: 2

**Conflict of interest disclosure statement**: None

**Declarations:**

Ethics approval and consent to participate: All methods were carried out in accordance with relevant guidelines and regulations in the declaration.

Consent for publication: All authors have agreed.

Availability of data and materials: Not applicable.

Competing interests: Not applicable.

Authors' contributions: MK have taken responsibility for everything.

Acknowledgements: The author thanks Associate Professor Hisashi Noma for his encouragement and helpful suggestions.

Patients are directly involved in this study: Not applicable.

**150-word statement of translational relevance:**

Recently, the strategy for dose optimization in oncology has shifted to conduct Phase 2 randomized controlled trials with multiple doses. Optimal biologic dose selection from Phase 1 trial data to determine candidate doses for Phase 2 trials has been gaining attention. This study proposes a novel adaptive cohort size determination method for optimal biologic dose-finding to accelerate trials. The cohort size expansion is determined adaptively depending on the toxicity and efficacy data of the ongoing trial. In a simulation, the proposed method shortened the trial duration and maintained accuracy. The trial duration was reduced by an average of approximately 20% compared with the non-adaptive cohort size determination design. The cohort size expansion is demonstrated using a simple example.

## Abstract

**Purpose:** Recently, the strategy for dose optimization in oncology has shifted to conduct Phase 2 randomized controlled trials with multiple doses. Optimal biologic dose (OBD) selection from Phase 1 trial data to determine candidate doses for Phase 2 trials has been gaining attention. Trials to identify the OBD have a fixed cohort size, which increases the trial duration. We propose a method to increase the cohort size using trial data and shorten the trial duration while maintaining accuracy.

**Methods**: We propose a novel adaptive cohort size determination method in which the increase of cohort size is determined using desirability probability on the basis of toxicity and efficacy data. The desirability probability is a measure of how desirable a dose is and thus how close it is to the OBD. However, in trial, the desirability probability does not need to be calculated. Instead, the cohort size expansion can be determined by a simple table generated in advance from toxicity and efficacy data. An illustrated example is provided and the performance was evaluated in a simulation study with 16 scenarios.

**Results**: In the simulation study, the trial duration was reduced by an average of 20% compared with the conventional design. The percentages of correct OBD selection are almost the same as those with the conventional design.

**Conclusion**: The proposed adaptive cohort size determination method described in this study reduces trial duration while maintaining accuracy.

## Introduction

Molecularly targeted agents and immunotherapy are highly effective approaches for the treatment of various cancers. While conventional cytotoxic chemotherapy shows dose-dependent increases in toxicity and efficacy, recent research has demonstrated that molecularly targeted agents and immunotherapy show different trends. The Dose Selection Strategies in Oncology workshop[1] in May 2022 presented dose-effect curves for cytotoxic chemotherapy vs. molecularly targeted agents vs. immunotherapy (approximately 18:30 min into the video of day 1). The results demonstrated that molecularly targeted agents increase efficacy up to a certain dose and then the efficacy curve becomes almost constant over a certain dose. Toxicity may continue to increase in a dose-dependent manner. Therefore, the maximum tolerated dose (MTD) that may be selected for Phase 2 trials might in fact have similar efficacy to that of a lower dose, in which the maximum effect has already been achieved but the toxicity rate is high. Shah et al.[2] reported that the MTD is frequently chosen for Phase 2 dose trials.

Immunotherapy may be able to exhibit efficacy at low doses; even if the dose is increased, efficacy may remain almost the same and toxicity may not change. Therefore, the methods for assessing anticancer drug toxicity and efficacy profiles in selecting MTD and recommended Phase 2 doses are problematic, and a different approach is required.

The FDA's Project Optimus recently described the optimal biologic dose (OBD). Shah et al.[2], the Friends of Cancer Research White paper[3], and FDA guidelines[4] stated that two or more doses should be selected from information obtained from Phase 1 trials and applied in Phase 2 randomized controlled trials. For selecting the two or more doses, it is preferable to collect more information (clinical pharmacokinetic, pharmacodynamic, and pharmacogenomic data) in the Phase 1 trial. Previous Phase 1 designs to identify the MTD have had the problem of insufficient dosing at doses lower than the MTD because dosing is concentrated near the MTD. Shah et al.[2] stated that "The answer to the dose-selection conundrum may sometimes be that less is more." OBD-identifying clinical trials are important because they actively search for the OBD in the dose-assignment.

The management of transformed chronic myeloid leukemia: ponatinib and intensive chemotherapy (MATCHPOINT) dose-finding study[5], conducted by the University of Birmingham, was a clinical trial that identified an OBD. Dose adjustment was performed using toxicity and efficacy data with the efficacy-toxicity trade-offs (efftox) model[6] to identify the OBD. While the efftox model is a design that can identify the OBD with high accuracy, physicians may feel that dose adjustment is obtained from information output from a black box, as the results from a model analysis are used when making dose adjustment decisions. Additionally, because of the complexity of the statistical model, the validity of the model needs to be examined in advance. Therefore, simple and easy-to-understand designs have been proposed. Model-assisted designs (U-BOIN[7], BOIN12[8], TITE-BOIN12[9], and BOIN-ET[10]) with fewer assumptions have been proposed as designs that explore OBD on the basis of safety and efficacy. Of these, the BOIN12 design is the simplest, because it also combines clinical interest in the BOIN design and efficacy. Lin et al.[8] compared the BOIN12 design with the efftox model and confirmed that the BOIN12 design is comparable to the efftox model. The BOIN12 design has been applied in a clinical trial at Beijing Boren Hospital (Clinical-Trials.gov identifier: NCT04835519)[11]. The study period tends to be long because of the time required to confirm efficacy data for each cohort. The TITE-BOIN12 design requires constant follow-up and dose adjustment of each cohort, which makes trial control very complicated when clinicians and related members are also involved in the trial. Philips and Mondal[12] showed that complex trial designs are often avoided. Moreover, Kakizume and Morita[13] researched a cohort size adaptation for a continual reassessment method design in which only safety was evaluated. This method is only applicable to a continual reassessment method design, which is complicated. Therefore, it is important to find a simpler method to shorten the trial time.

This study proposes a method to accelerate the trial by adjusting the cohort size for the next cohort using the trial toxicity and efficacy data. If the dose of next cohort is likely to be the OBD, the cohort size will be greater than three; if the number of toxicities is high and the number of patients that obtain treatment efficacy is low, the cohort size is maintained at three as usual. Because the number of patients with toxicity and the number of patients who obtain treatment efficacy along with cohort size can be presented in advance in a simple table, the simplicity of the BOIN12 design is maintained. An example study is provided here to illustrate the process of cohort size adjustment. Computer simulations were used to evaluate the performance of the proposed design.

## Materials and Methods

The BOIN12 design first divides the outcomes of interest into four categories. The first is "no toxicity and efficacy," the second is "toxicity and efficacy," the third is "no toxicity and no efficacy," and the last is "toxicity and no efficacy." Each outcome is given a desirability score. The most desirable outcome, "no toxicity and efficacy," is given a score of 100; the next most desirable, "toxicity and efficacy," is given a score of 60; the third most desirable, "no toxicity and no efficacy," is given a score of 40; and the undesirable outcome, "toxicity and no efficacy," is given a score of 0. From the weighted sum of scores and the probabilities of the four outcomes, the BOIN12 design translates toxicity and efficacy data into a single index of desirability value, ranging from 0 to 100. The probability that the desirability value exceeds the highest undesirable value is calculated; the numbers are assigned in ascending order from 1, starting with the toxicity and efficacy combination with the lowest probability. The assigned number is used as the desirability score, and the dose with the highest desirability score is selected as the next dose during dose adjustment. For an example of dose-assignment with the BOIN12 design, the dose-assignment flow chart is shown in Figure 1 and the dose-assignment decision tables are shown in Table 1. For a trial with a sample size of 18, when three patients were treated at dose 1 and (the number of patients with toxicity, the number of patients that achieved treatment efficacy)=(0,1), the desirability score of dose 1 is 255. Dose 2 has not yet been administered, but a desirability score was calculated at 269. We use the desirability probability to adaptively determine the cohort size. The desirability probability is close to 1 when the effectiveness is high and the toxicity fraction is low, and it is close to 0 when the effectiveness is low and the toxicity fraction is high. In other words, the closer the desirability probability is to 1, the more suitable the dose is for the OBD. The technical details for desirability probability are given in Supplemental Description S1. The cohort size is increased when the desirability probability exceeds a predetermined threshold. If the threshold is set at 0.2, for example, the cohort size is increased to 6 when the conditions in Table 2 are met. Under the conditions of the BOIN12 design above, the recommended value of the threshold is 0.20. This is because the cohort size can be increased if 1/3 of the efficacy is obtained in the absence of toxicity in up to 9 patients treated. The cohort size of next cohort is increased if the next cohort dose met the conditions on Table 2. An illustrative example is provided below.

**Illustrative example**

We consider a virtual dose-finding trial with a sample size of 18.

[Example 1] The increase in cohort size is illustrated in Example 1 (Figure 2). Dose 1 is administered to three patients in cohort 1. After completion of the safety and efficacy evaluation, following the dose-assignment rules in Table 1 and Figure 1, dose 2 will be the dose for the next cohort. For the adaptive cohort size determination BOIN12 (AD-BOIN12) design, dose 2 does not meet the conditions in Table 2. Hence, the cohort size is maintained. After completion of the evaluation of cohort 2, the dose for the next cohort will be dose 2, as shown in Table 1 and Figure 1. For the AD-BOIN12 design, because dose 2 satisfies the conditions in Table 2, the cohort size is increased to six. For the BOIN12 design, the dose assignment is repeated in the same procedure until the sample size reaches the maximum sample size. For the AD-BOIN12 design, dose 1 is selected after the evaluation of cohort 3 and Table 2, resulting in a cohort size of six. Adjusting the cohort size in the AD-BOIN12 design reduced the study duration by 100 days.

[Example 2] The results for cohort 1 dosing were the same as in Example 1. After cohort 2 evaluation, dose 3 was selected for the next cohort. After evaluation of cohort 3, dose 2 was selected for the next cohort. For the AD-BOIN12 design, because dose 2 satisfies the conditions in Table 2, the size of the next cohort is six; after the evaluation of cohort 4, dose 2 is still selected, which satisfies the conditions in Table 2, but the size of the next cohort is 3 because increasing the cohort size will exceed the maximum sample size of 18. The AD-BOIN12 design shortened the trial by 50 days compared with the BOIN12 design.

## Numerical simulation study

We used a simulation study to evaluate the performance of the adaptive cohort size determination BOIN12 (AD-BOIN12) design compared with the BOIN12 design. We used the simulation setup as described by Lin et al[8]. We prepared 16 simulation scenarios of toxicity probabilities and efficacy probabilities, shown in Supplemental Table S1. We assume that the sample size is 36 and the dose level is 5. The utility scores of (efficacy and no toxicity, efficacy and toxicity, no efficacy and no toxicity, and no efficacy and toxicity) were (100, 60, 40, 0). The accrual rate was three patients per month. The toxicity assessment period was 45 days. The efficacy assessment period was 60 days. In Case A, we applied the stopping rule, in which the trial was completed when 12 patients were treated at the current dose and the dose of the next cohort is the same as the current dose. In Case B, the stopping rule was not applied. The number of simulation times was 10,000. The maximum acceptable toxicity probability was 0.35. The minimum acceptable efficacy probability was 0.25. As a sensitivity analysis, we prepared a case with a changed toxicity and efficacy assessment period. Case C has a toxicity assessment period of 45 days and an efficacy assessment period of 30 days (with the stopping rule). Case D has a toxicity assessment period of 45 days and efficacy assessment period of 30 days (no stopping rule). The cutoff in safety admissibility criteria was set at 0.95. The cutoff in efficacy admissibility criteria was set 0.90. When there was no admissible dose in dose-assignment, the trial was stopped early for all simulations. As a supplemental analysis, we used the TITE-BOIN12 design. Because of the complexity of the TITE-BOIN12 design, the supplemental simulation study was implemented by trialdesign.org[14]. The BOIN12 and AD-BOIN12 designs were simulated using R software. The simulation program files are included in the Supplemental material. The evaluation criteria were as follows.

**Evaluation criteria**

1. Percentage of correct OBD selection
2. Trial duration (months)
3. Number of patients treated at the correct OBD
4. Number of patients treated at toxic doses

Data Availability: Not applicable.

## Results

A simulation study was performed to evaluate the performance of the AD-BOIN12 design using the adaptive cohort size determination compared with the BOIN12 design. A total of 16 simulation scenarios were prepared. The results for Case A are shown in Figure 3 and those for Case B are shown in Figure 4.

**Performance of correct OBD selection.** The percentage of the correct OBD selection for 14 scenarios, excluding scenario 15 and 16 without OBD, are shown in Figure 3A and 4A. For Case A, the average difference of the percentage of the correct OBD selection of the AD-BOIN12 and BOIN12 designs (AD-BOIN12 - BOIN12) was -0.3%, the minimum difference was –1.7%, and the maximum difference was 1.5%. In Case B, the average difference between the percentage of the correct OBD selection was -0.8%, the minimum difference was –2.6%, and the maximum difference was 2.4%. The detailed results of each scenario and each Case are shown in Supplemental Table S1 and S2. The results of evaluation windows changed for safety and efficacy are included in Supplemental Table S3 and S4 and Supplemental Figure S1 and S2.

**Trial duration (months).** In Case A, the AD-BOIN12 design reduced the trial duration by an overage of 3.2 months compared with BOIN12 design, with a minimum reduction of 2.3 months and a maximum reduction of 4.1 months. The average reduction in the BOIN12 design with the trial duration as baseline was 14.2%. For Case B, the AD-BOIN12 design reduced the trial duration by an average of 5.7 months compared with the BOIN12 design, with a minimum reduction of 3.1 months and a maximum reduction of 7.9 months. The average reduction in the BOIN12 design with the trial duration as baseline was 19.0%. Scenarios without OBD were found to reduce the trial duration.

**Number of patients treated at the correct OBD.** For Case A, the average difference in the number of patients treated at the correct OBD was 0.3, the minimum difference was –0.5, and the maximum difference was 0.7. For Case B, the average difference of the number of patients treated at the correct OBD was -0.8, the minimum difference was –2.6, and the maximum difference was 2.4.

**Number of patients treated at toxic doses.** For Case A, the average difference in the number of patients treated at toxic doses was 0.0, the minimum difference was –0.2, and the maximum difference was 0.3. For Case B, the average difference in the number of patients treated at toxic doses was -0.1, the minimum difference was –0.4, and the maximum difference was 0.1.

## Discussion

This study proposes a novel adaptive cohort size determination BOIN12 design to accelerate dose-finding trials. For doses that are effective, as determined by data from clinical trials, the cohort size should be increased. We confirmed that the AD-BOIN12 design does not decrease accuracy compared with the BOIN12 design.

In this study, we illustrated how to increase cohort size using two practical examples. For a virtual clinical trial with a sample size of 18 cases, the AD-BOIN12 design was shown to shorten the study period by 50–100 days. The simulation study evaluated the performance of the AD-BOIN12 design. The percentages of correct OBD selection in all scenarios were almost same as those with the BOIN12 design. However, for Case A, in scenarios 4, 11, 12, and 14, the percentages of correct OBD selection were slightly higher for the AD-BOIN12 design. For Case B, in scenarios 4, 11, and 14, the percentage of correct OBD selection were slightly higher for the AD-BOIN12 design. We confirmed that the trial duration could be reduced from any scenario with the AD-BOIN12 design compared with the BOIN12 design. We confirmed that the TITE-BOIN12 design in the supplemental analyses can shorten the study period compared with the BOIN12 and AD-BOIN12 designs. However, our proposed method is more convenient because the cohort size can be adjusted using information from a simple table. For Case A and Case B, the numbers of patients treated at the correct OBD in all scenarios were almost same as those with the BOIN design. However, for Case A, the number of patients treated increased slightly in all scenarios except for scenario 3 and 9. For Case B, the number of patients treated increased slightly in scenarios 2, 5, 8, 9, 11, 12, and 13. For Case A and Case B, the numbers of patients treated at the correct OBD in all scenarios were almost same as those with the BOIN12 design. Increasing the cohort size did not increase the number of patients treated at toxic doses. Similar trends were observed in case C and case D.

We confirmed that the performance of the AD-BOIN12 designs was acceptable. Because the study results showed nearly the same performance of the AD-BOIN12 and BOIN12 designs, we recommend using the AD-BOIN12 design. As the need for OBD identification grows, we expect AD-BOIN12 to play an increasingly active role in the future.

**Acknowledgments**: The author thanks Associate Professor Hisashi Noma for his encouragement and helpful suggestions.

## Author's Contributions

M. Kojima: Conception and design; development of methodology; acquisition of data (e.g., provided animals, acquired and managed patients, provided facilities); analysis and interpretation of data (e.g., statistical analysis, biostatistics, computational analysis); writing, review, and revision of the manuscript; administrative, technical, and material support (i.e., reporting and organizing data, constructing databases); and study supervision.

**Table 1. Dose-assignment decision tables**

For safety

| Patients treated (n) | 3 | 6 | 9 | 12 | 15 |
|---|---|---|---|---|---|
| d(n): Escalate if # of patients with toxicity ≤ | 0 | 1 | 2 | 3 | 4 |
| e(n): De-escalate if # of patients with toxicity ≤ | 2 | 3 | 4 | 6 | 7 |

For desirability score

| Patients (n) | No Tox | No Eff | DS |
|---|---|---|---|
| 0 | 0 | 0 | 269 |
| 3 | 0 | 0 | 186 |
| 3 | 0 | 1 | 255 |
| 3 | 0 | 2 | 326 |
| 3 | 2 | 1 | 164 |
| 6 | 0 | 3 | 296 |
| 6 | 1 | 2 | 208 |
| 9 | 1 | 3 | 202 |
| 9 | 1 | 4 | 245 |
| 12 | 1 | 4 | 197 |

DS: desirability score. No Tox: number of patients with toxicity. No Eff: number of patients with efficacy.

This table presents only the information necessary for the dose assignment in this study.

**Table 2. Adaptive cohort size determination**

Cohort size when desirability probability > 0.20

| Patients (n) | No Tox | No Eff |
|---|---|---|
| 3 | 0 | ≥1 |
| 3 | 1 | ≥2 |
| 6 | 0 | ≥2 |
| 6 | 1 | ≥3 |
| 6 | 2 | ≥4 |
| 9 | 0 | ≥3 |
| 9 | 1 | ≥4 |
| 9 | 2 | ≥5 |
| 9 | 3 | ≥5 |

No Tox: number of patients with toxicity. No Eff: number of patients with efficacy.

Cohort size when desirability probability > 0.25

| **Patients (n)** | No Tox | No Eff |
|---|---|---|
| 3 | 0 | ≥1 |
| 3 | 1 | ≥2 |
| 6 | 0 | ≥3 |
| 6 | 1 | ≥3 |
| 6 | 2 | ≥4 |
| 9 | 0 | ≥4 |
| 9 | 1 | ≥5 |
| 9 | 2 | ≥5 |
| 9 | 3 | ≥6 |

No Tox: number of patients with toxicity. No Eff: number of patients with efficacy.

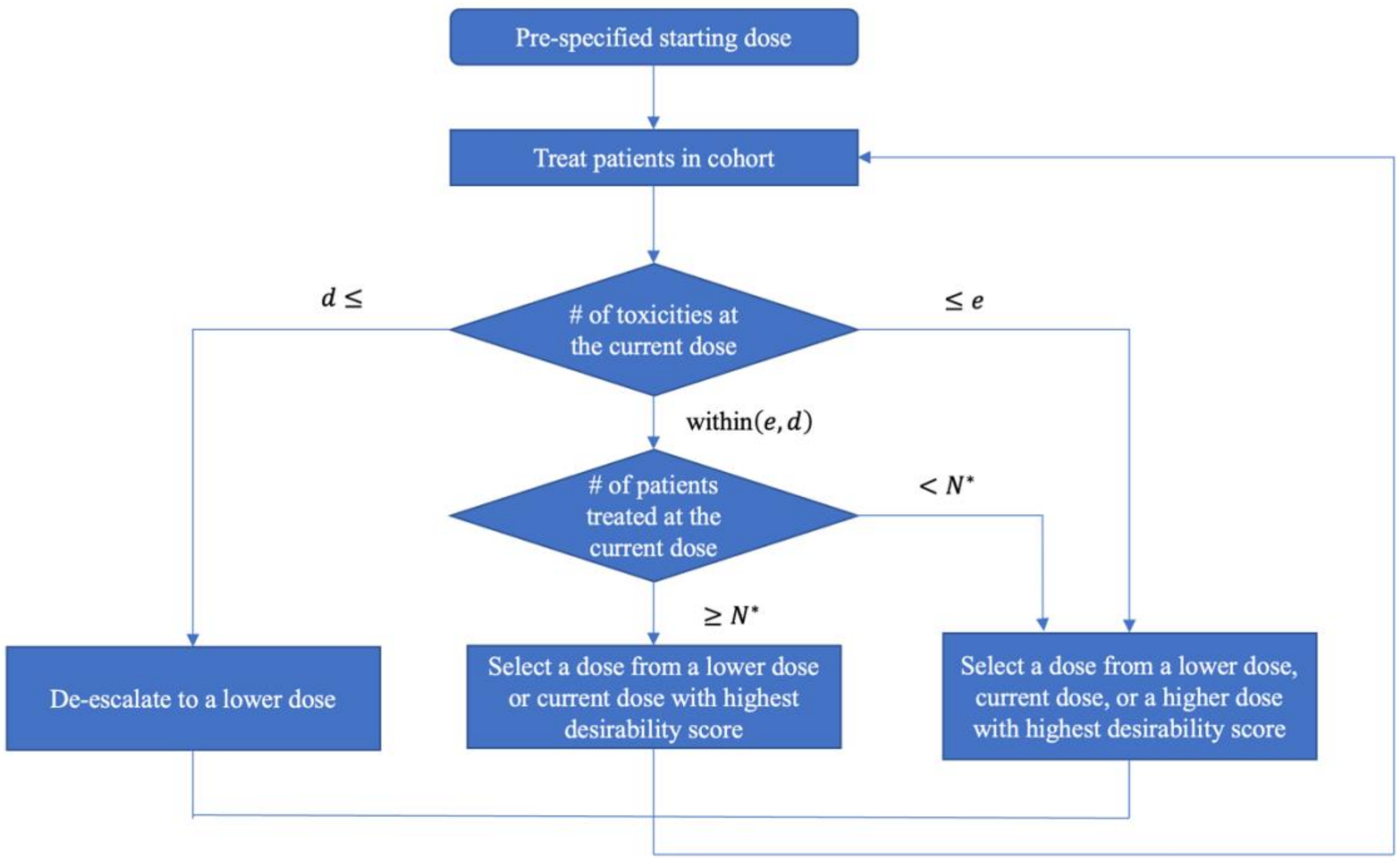

**Figure 1.** Dose-assignment flow chart.

[A]

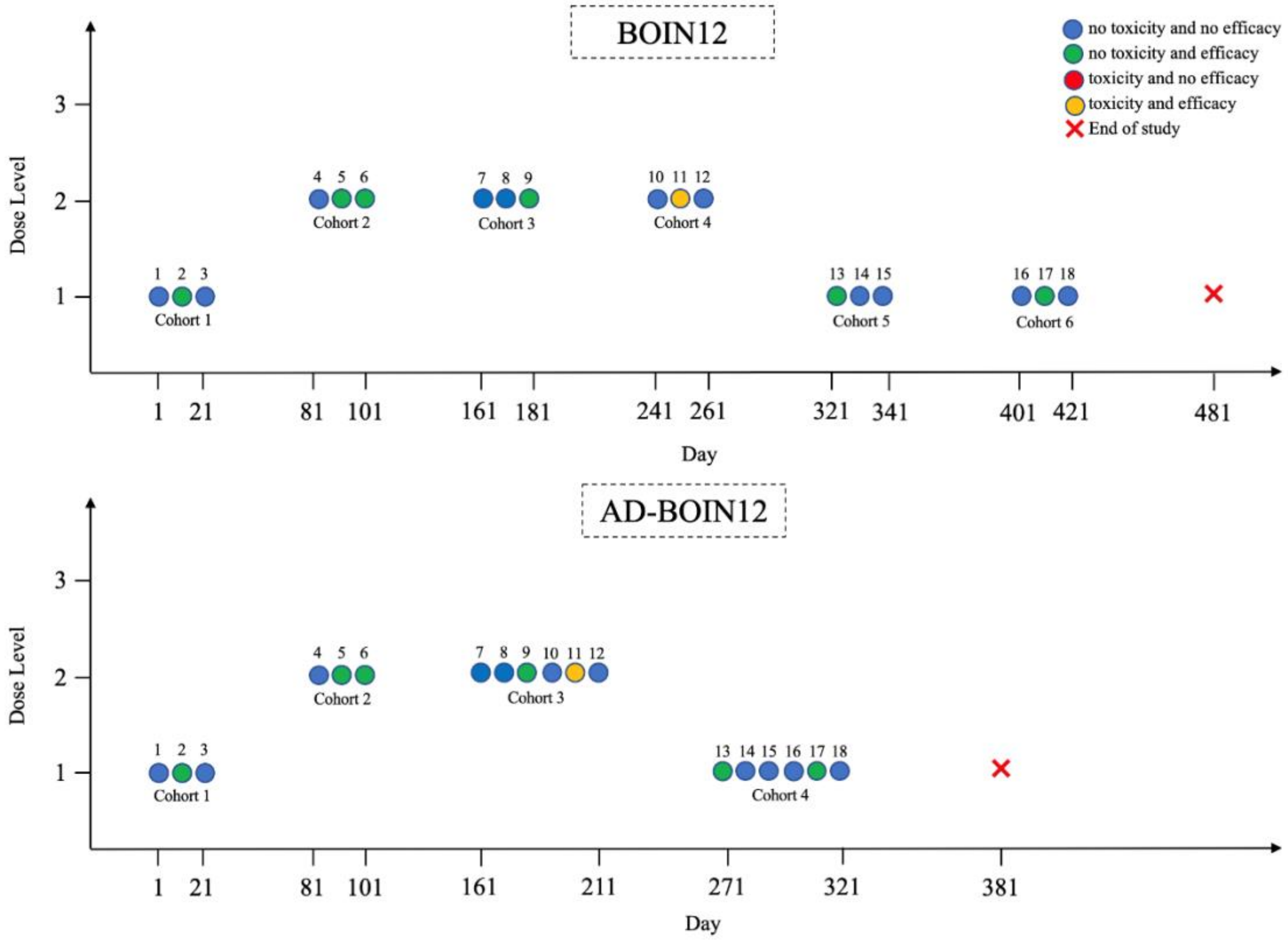
BOIN12
no toxicity and no efficacy
no toxicity and efficacy
toxicity and no efficacy
toxicity and efficacy
End of study
Dose Level
3
2
1
1 2 3
Cohort 1
4 5 6
Cohort 2
7 8 9
Cohort 3
10 11 12
Cohort 4
13 14 15
Cohort 5
16 17 18
Cohort 6
1 21 81 101 161 181 241 261 321 341 401 421 481
Day
AD-BOIN12
Dose Level
3
2
1
1 2 3
Cohort 1
4 5 6
Cohort 2
7 8 9 10 11 12
Cohort 3
13 14 15 16 17 18
Cohort 4
1 21 81 101 161 211 271 321 381
Day

[B]

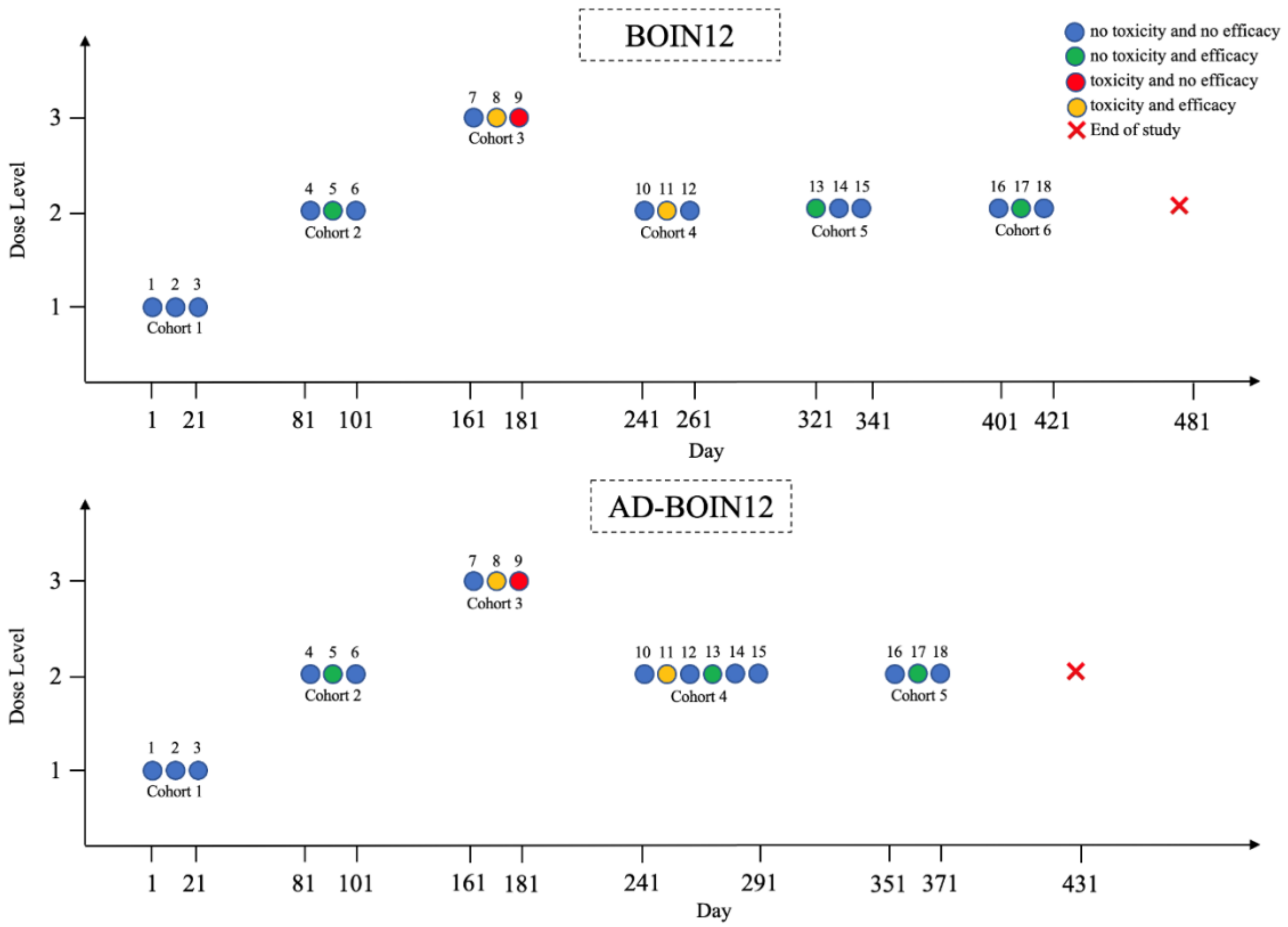

**Figure 2. Example of dose-assignment** (A) Example 1. (B) Example 2.
AD-BOIN12: Adaptive cohort size determination BOIN12

[A]

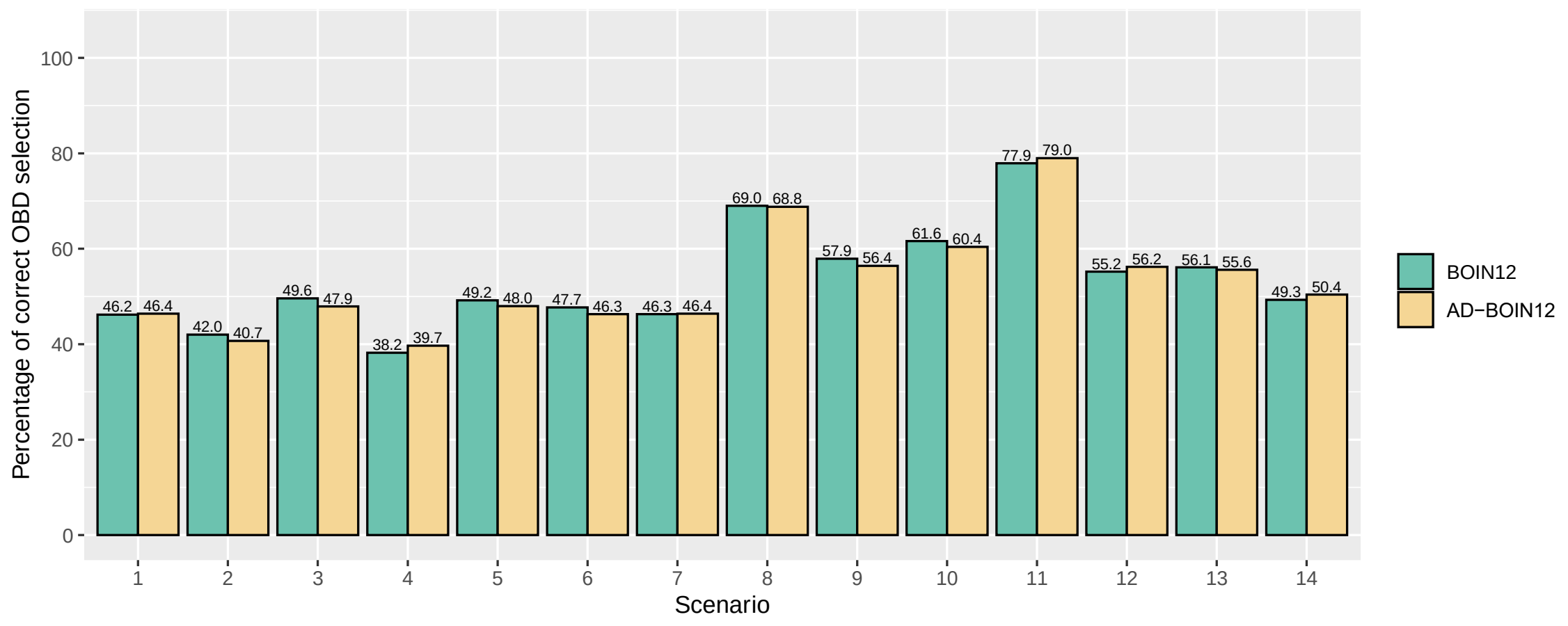

[B]

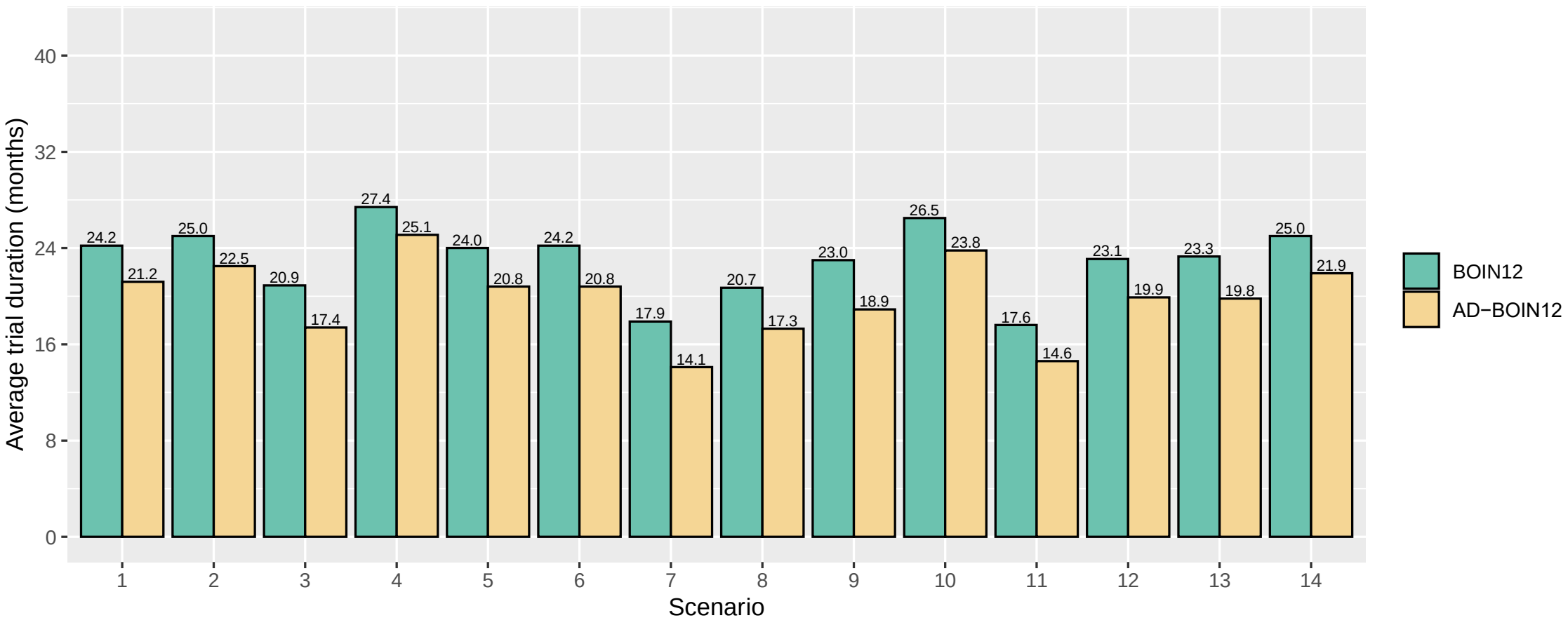

[C]

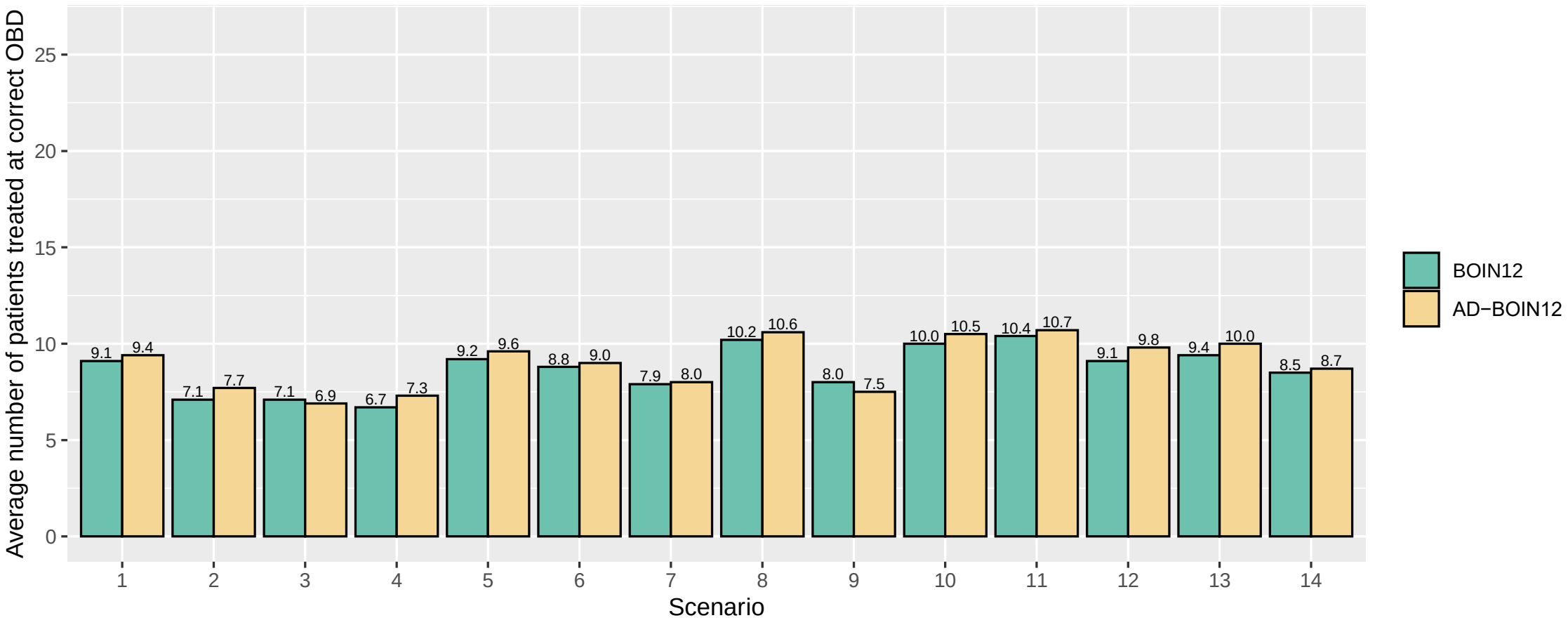

Average number of patients treated at correct OBD
0
5
10
15
20
25
9.1
9.4
7.1
7.7
7.1
6.9
6.7
7.3
9.2
9.6
8.8
9.0
7.9
8.0
10.2
10.6
8.0
7.5
10.0
10.5
10.4
10.7
9.1
9.8
9.4
10.0
8.5
8.7
1
2
3
4
5
6
7
8
9
10
11
12
13
14
Scenario
BOIN12
AD−BOIN12

[D]

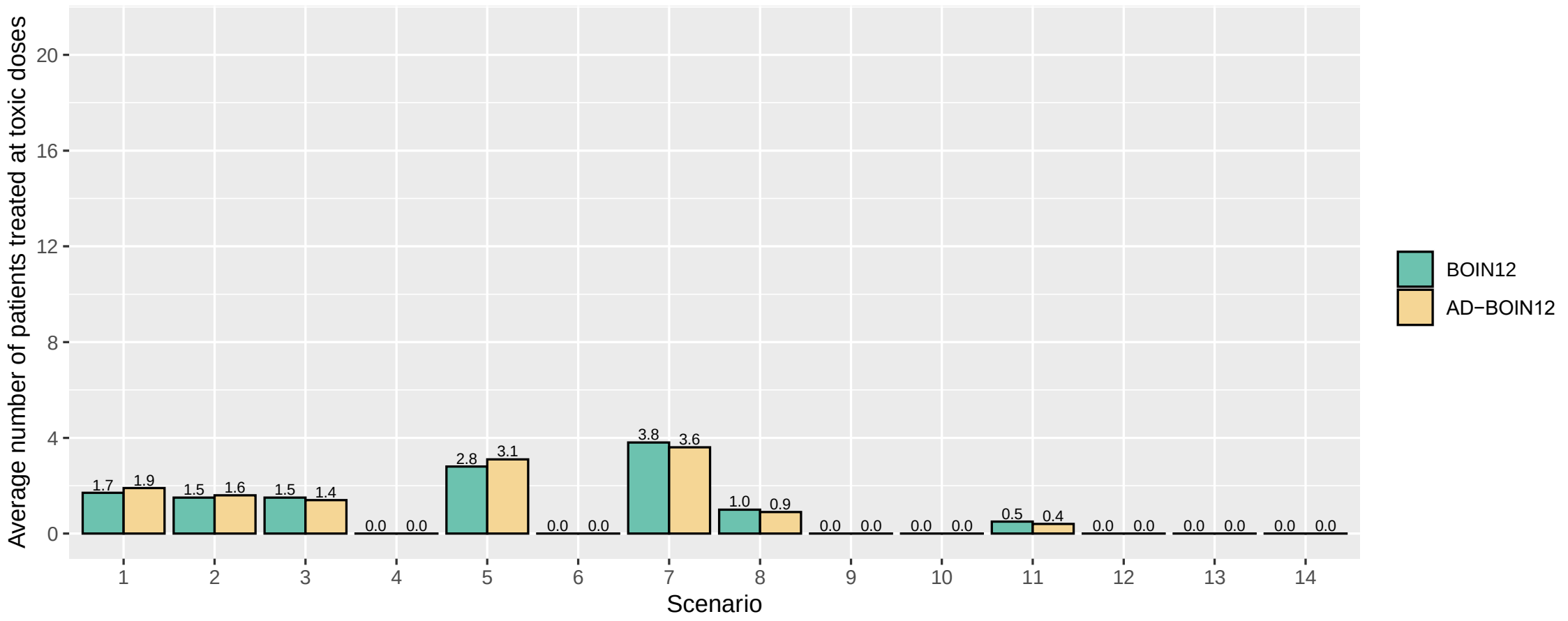

**Figure 3.** Simulation results for Case A. (A) Percentage of correct OBD selection. (B) Average trial duration (months). (C) Average number of patients treated at correct OBD. (D) Number of patients treated at toxic doses.

AD-BOIN12: Adaptive cohort size determination BOIN12

[A]

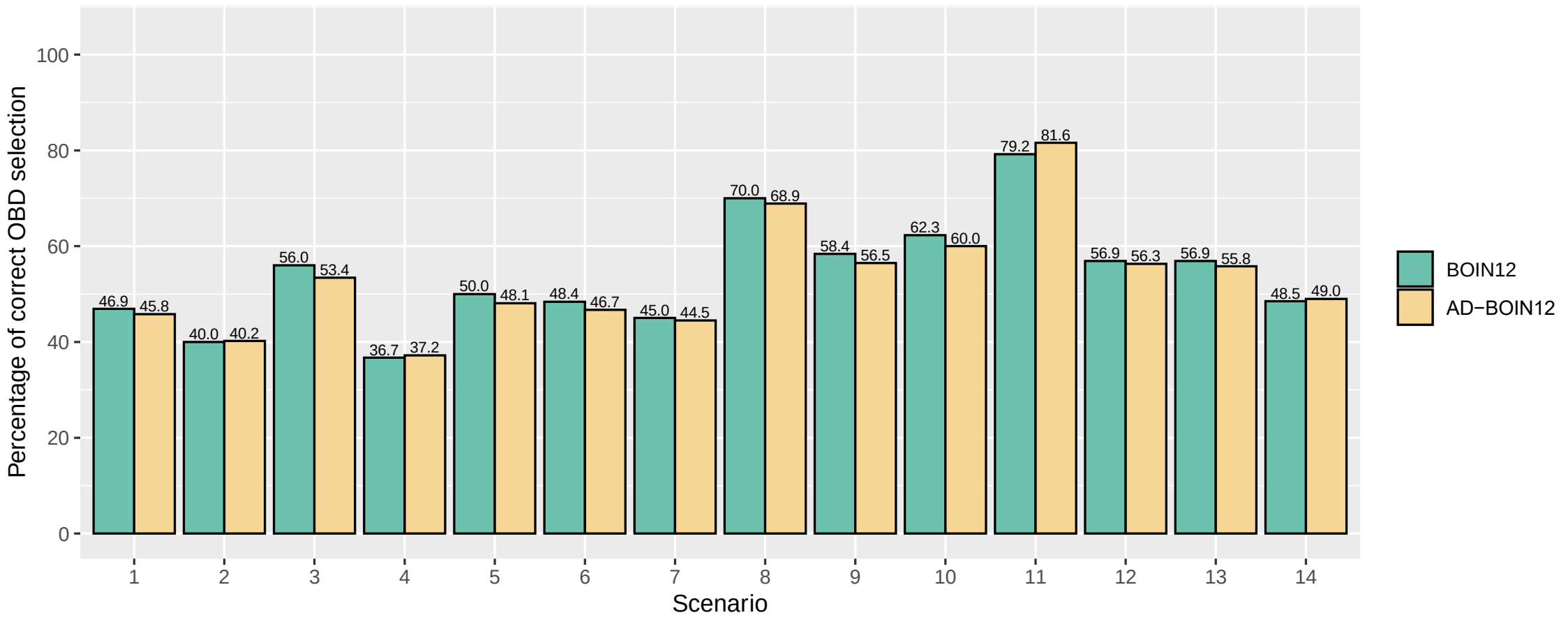

[B]

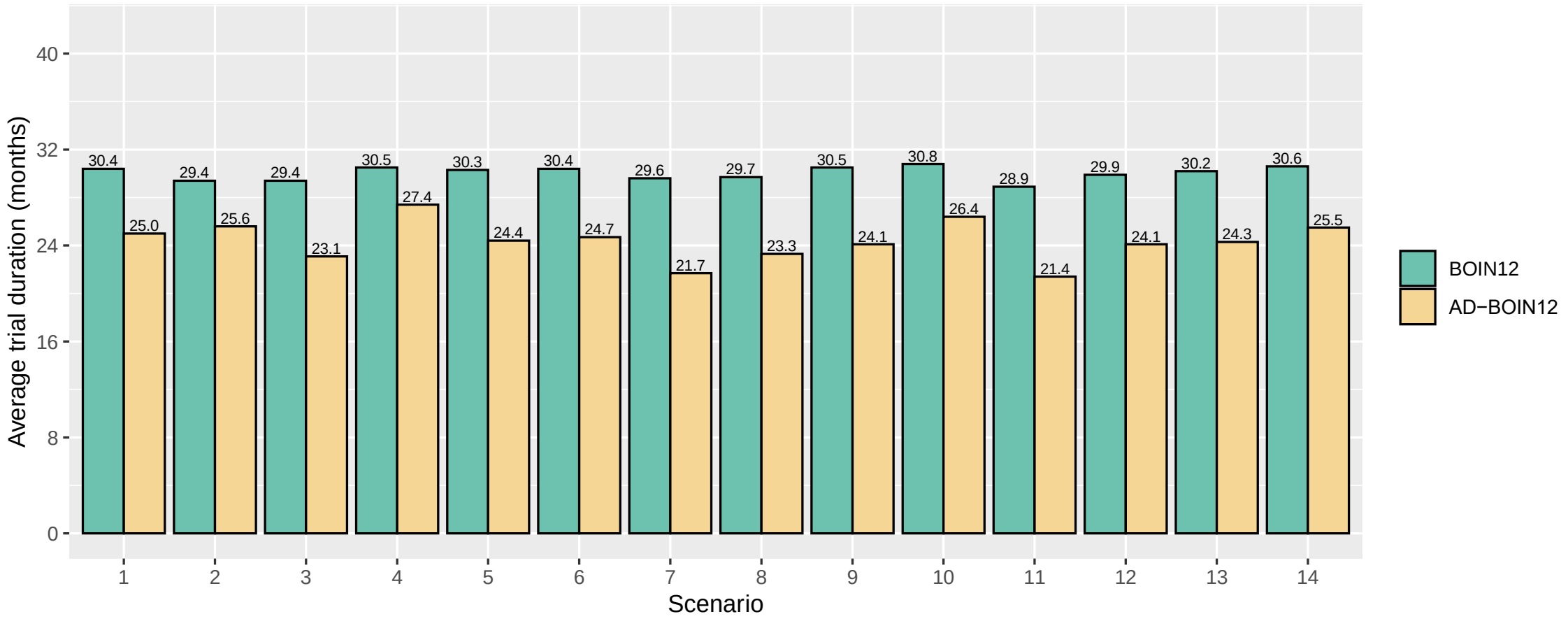
Average trial duration (months)
Scenario
BOIN12
AD−BOIN12
30.4 25.0 29.4 25.6 29.4 23.1 30.5 27.4 30.3 24.4 30.4 24.7 29.6 21.7 29.7 23.3 30.5 24.1 30.8 26.4 28.9 21.4 29.9 24.1 30.2 24.3 30.6 25.5

[C]

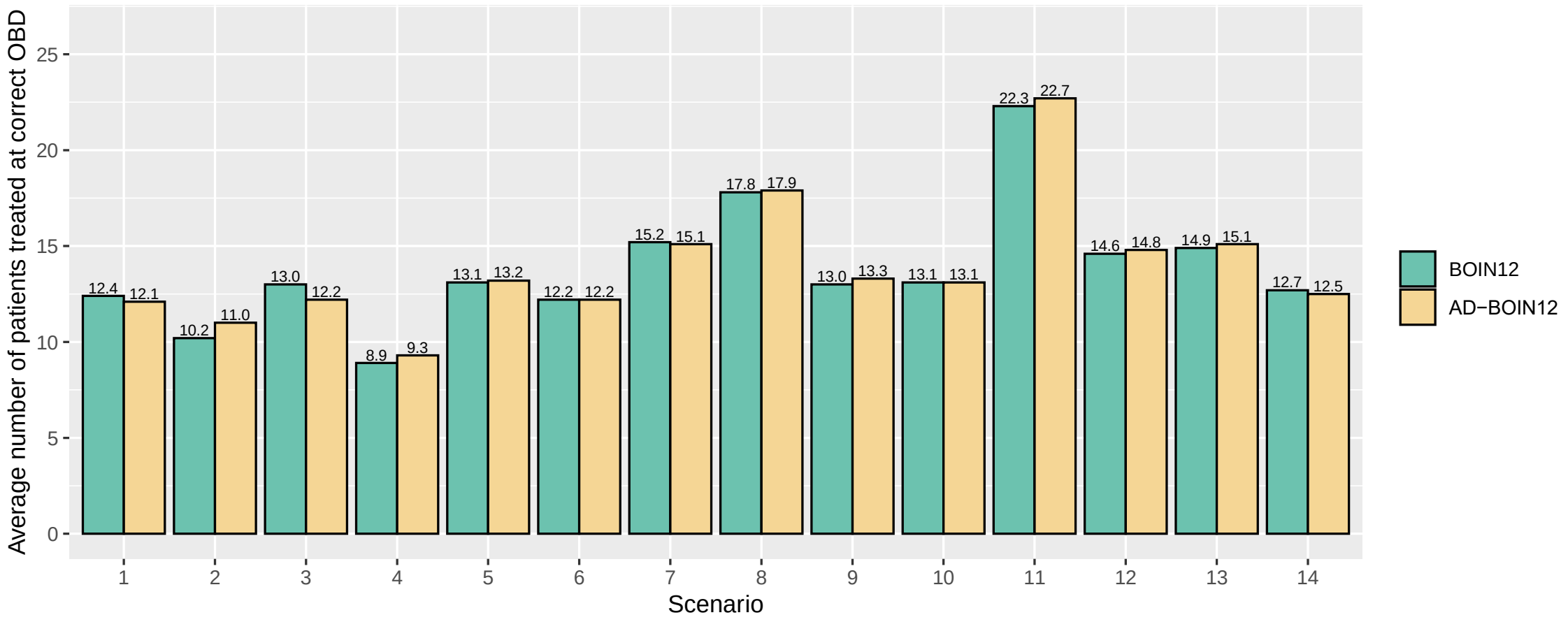
Average number of patients treated at correct OBD
Scenario
BOIN12
AD−BOIN12
12.4 12.1 10.2 11.0 13.0 12.2 8.9 9.3 13.1 13.2 12.2 12.2 15.2 15.1 17.8 17.9 13.0 13.3 13.1 13.1 22.3 22.7 14.6 14.8 14.9 15.1 12.7 12.5

[D]

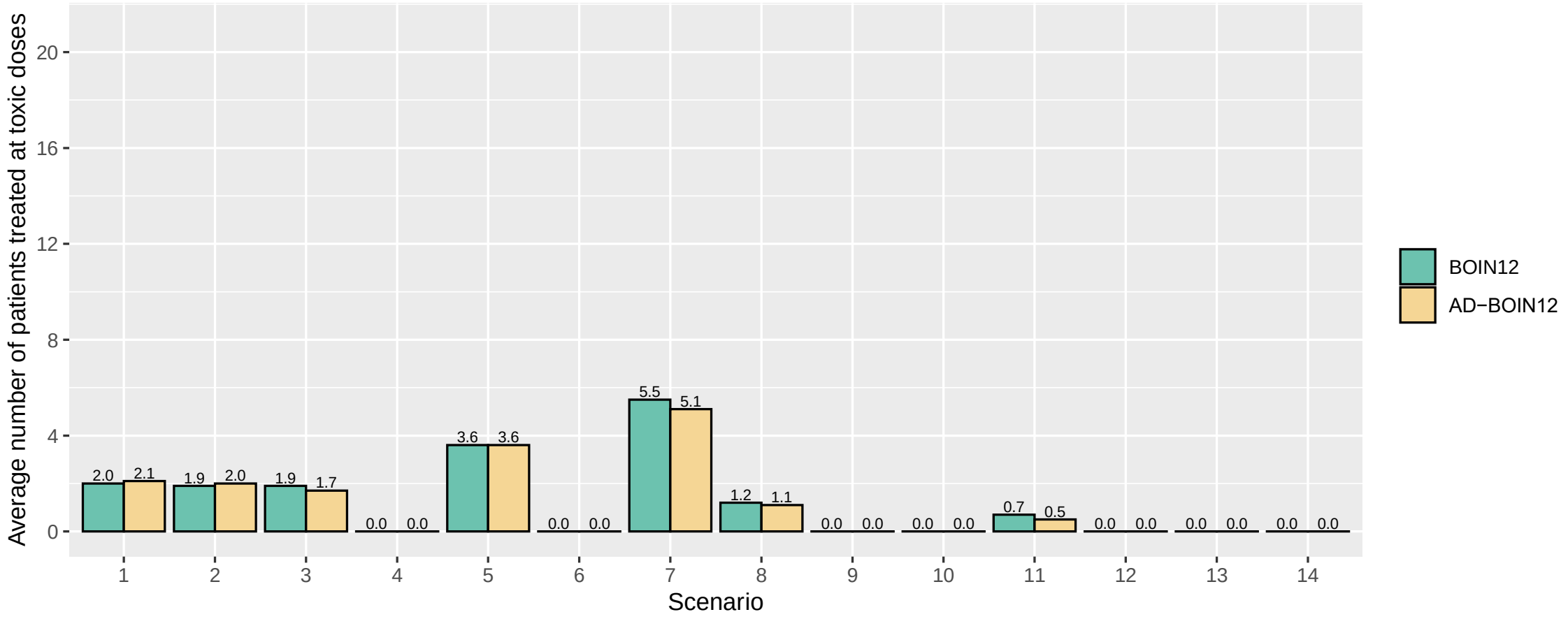
Average number of patients treated at toxic doses
Scenario
BOIN12
AD−BOIN12
2.0 2.1 1.9 2.0 1.9 1.7 0.0 0.0 3.6 3.6 0.0 0.0 5.5 5.1 1.2 1.1 0.0 0.0 0.0 0.0 0.7 0.5 0.0 0.0 0.0 0.0 0.0 0.0

**Figure 4.** Simulation results for Case B. (A) Percentage of correct OBD selection. (B) Average trial duration (months). (C) Average number of patients treated at correct OBD. (D) Number of patients treated at toxic doses.

AD-BOIN12: Adaptive cohort size determination BOIN12